\documentclass[journal]{IEEEtran}
\IEEEoverridecommandlockouts
\usepackage{cite}
\usepackage{amsmath,amssymb,amsfonts}
\usepackage[cmintegrals]{newtxmath}
\usepackage{graphicx}
\usepackage{textcomp}
\usepackage{xcolor}
\usepackage{multirow}
\usepackage{orcidlink}
\makeatletter
\let\NAT@parse\undefined
\makeatother
\usepackage{hyperref}

\begin{document}

\title{Learning of Time-Frequency Attention Mechanism for Automatic Modulation Recognition}

\author{
    Shangao~Lin\orcidlink{0000-0002-0850-7696}, \IEEEmembership{Student~Member,~IEEE},
    Yuan~Zeng\orcidlink{0000-0002-5260-8064}, \IEEEmembership{Member,~IEEE},
    and~Yi~Gong\orcidlink{0000-0001-7392-8991}, \IEEEmembership{Senior~Member,~IEEE}
    \thanks{
        Manuscript received Month Day, Year; revised Month Day, Year; accepted Month Day, Year. Date of publication Month Day, Year; date of current version Month Day, Year.
        This work is supported by National Key R\&D Program of China under Grant 2019YFB1802800, National Natural Science Foundation in China under Grants 62071212 and 62106095, Guangdong Science and Technology Program under Grant 2019A1515110479, Guangdong Basic and Applied Basic Research Foundation under Grant 2019B1515130003, Shenzhen Science and Technology Program under Grant JCYJ202001091414409, and Educational Commission of Guangdong Province of China under Grants 2020ZDZX3057, 2019KQNCX128 and 2017KZDXM075.
        The editor coordinating the review of this article and approving it for publication was C.-K. Wen.
        \emph{(Corresponding authors: Yuan Zeng and Yi Gong.)}

        S. Lin and Y. Gong are with the Department of Electrical and Electronic Engineering, Southern University of Science and Technology, Shenzhen 518055, China (e-mail: gongy@sustech.edu.cn).

        Y. Zeng is with the Academy for Advanced Interdisciplinary Studies, Southern University of Science and Technology, Shenzhen 518055, China (e-mail: zengy3@sustech.edu.cn).

        Digital Object Identifier 10.1109/LWC.2021.XXXXXXX}
}

\maketitle

\begin{abstract}
    Recently, deep learning-based image classification and speech recognition approaches have made extensive use of attention mechanisms to achieve state-of-the-art recognition, which demonstrates the effectiveness of attention mechanisms. Motivated by the fact that the frequency and time information of modulated radio signals are crucial for modulation recognition, this paper proposes a time-frequency attention mechanism for convolutional neural network (CNN)-based automatic modulation recognition. The proposed time-frequency attention mechanism is designed to learn which channel, frequency and time information is more meaningful in CNN for modulation recognition. We analyze the effectiveness of the proposed attention mechanism and evaluate the performance of the proposed models. Experiment results show that the proposed methods outperform existing learning-based methods and attention mechanisms.
\end{abstract}


\IEEEpeerreviewmaketitle

\section{Introduction}
\IEEEPARstart{A}{utomatic} modulation recognition (AMR) is the task of classifying the modulation mode of radio signals received from wireless communication systems.
It is an intermediate step between signal detection and signal demodulation. As a step towards understanding what type of communication scheme and emitter is present, AMR has been widely used in practical civilian and military applications, such as cognitive radio, spectrum monitoring, communications interference, and electronic countermeasures.

In the past few years, due to the great success in computer vision and natural language processing, data-driven deep learning methods have also been applied to AMR, showing great potential in improving recognition accuracy and robustness.
O'Shea \textit{et al.} generated an open modulation recognition dataset, called RadioML2016.10a, using GNU Radio in \cite{cite:radioML}, and first proposed a deep neural network (DNN) architecture for AMR in \cite{cite:IQCNN}. Later, various DNN architectures were introduced to improve the recognition accuracy, such as convolutional neural networks \cite{cite:li2018robust}, recurrent neural networks \cite{cite:rajendran2018deep}, and graph convolutional network \cite{cite:liu2020gcn}. Recently, a few deep learning-based approaches considered the inherent properties of radio signals and communication systems in modulation recognition. Yashashwi \textit{et al.} \cite{cite:yashashwi2019cm} proposed a learnable distortion correction module to shift the frequency and phase of signal according to its weights and jointly train with a CNN. In \cite{cite:zhang2018automatic}, high-order statistics of radio signal was computed as an additional signal representation to the CNN classifier. Zeng \textit{et al.} \cite{cite:SCNN} exploited the time-frequency analysis of modulated radio signals and proposed a CNN with the short-time discrete Fourier transform (STFT). Wang \textit{et al.} \cite{cite:wang2021cue} proposed a multi-cue fusion network by modelling spatial-temporal correlations from modulated signal cues. Our work further leverages time-frequency characteristics of time series during the design of attention mechanism for modulation recognition.

In addition to DNNs, attention mechanisms have also been used in a wide variety of DNN-based methods in computer vision. A neural attention module can optimize the weights of the input features by minimizing recognition errors. This can hence enhance the important information and reduce the interference caused by irrelevant information in learning-based recognition frameworks. In \cite{cite:SE-net}, a squeeze-and-excitation (SE) attention was proposed. It computes channel attention with the help of 2D global pooling and provides notable performance gains at a considerably low computational cost. In \cite{cite:CBAM}, Woo \textit{et al.} proposed a convolutional block attention module, which sequentially implements channel attention and spatial attention to enhance important parts of the input features. In contrast, our attention mechanism attends features in channel, frequency and time dimensions for improving the modulation recognition performance of existing CNN-based models.

In this work, we propose a time-frequency attention (TFA) mechanism to learn useful features from spectrogram images in terms of channel, frequency and time, and improve the recognition performance of existing methods.
The channel attention is performed first to learn weights regarding channel importance in the input feature map, and then frequency and time attention mechanisms are performed in parallel and composited using learned weights for capturing both frequency and time attention. In addition, we integrate the proposed TFA mechanism into two CNN-based AMR models to improve the performance of AMR. Moreover, we conduct ablation experiments to analyze the effectiveness of the proposed TFA mechanism, and compare the presented AMR models with three existing state-of-the-art (SOTA) AMR methods in \cite{cite:IQCNN,cite:CLDNN,cite:liang2021automatic}.

\section{Problem Statement and Spectrum Representation}
This paper considers a simple single-input single-output wireless communication system, where a symbol is converted and transmitted to a receiver via a communication channel. The data model of received signal $r(t)$ is given as:
\begin{equation}
    r(t)=\mathcal{F}(s(t)) * h(t)+n(t),
\end{equation}
where $s(t)$ denotes the transmission symbol, $\mathcal{F}$ is a modulation function, $h(t)$ is the communication channel impulse response, and $n(t)$ is the additive white Gaussian noise. Given the received signal $r(t)$, AMR aims at decoding the modulation function $\mathcal{F}$. A discrete-time version of the continuous-time signal $r(t)$ can be obtained by sampling $r(t)$ for $n$ times with a sampling rate $f_s=\frac{1}{T_s}$ i.e. $r(n)=r(t)|_{t=nT_s}, -\infty<n<+\infty$.

Since the time-frequency analysis of a modulated signal can reflect its frequency varies with time, which is an important distinction among different modulated signals. In this work, we exploit the insight from recent work \cite{cite:SCNN} that spectrograms can achieve richer time-frequency representation of signals, and use STFT-based spectrogram to represent the signal about frequency variation trend with time. The continuous signal $r(t)$ is first converted to discrete-time signal $r(n)$ with sampling frequency $f_s$, and
then $r(n)$ is windowed and transformed into the frequency domain by applying the STFT, that is:
\begin{equation}
    R(m, k)=\sum_{n=m K}^{m K+(L-1)} r(n) w(n-m K) e^{-j \frac{2 \pi k}{L} n},
\end{equation}
\noindent where $m$ and $k$ denote the time frame and frequency bin indices, respectively. $w(n)$ denotes the window function, $L$ is the frame length, and $K$ is the frame shift. The spectrogram $x$ is given as $x=|R(m, k)|^{2}$, where each pixel corresponds to a point in frequency and time.

\section{Automatic Modulation Recognition}
\subsection{Time-Frequency Attention}
\begin{figure}[htbp]
    \centering{\includegraphics[scale=0.25]{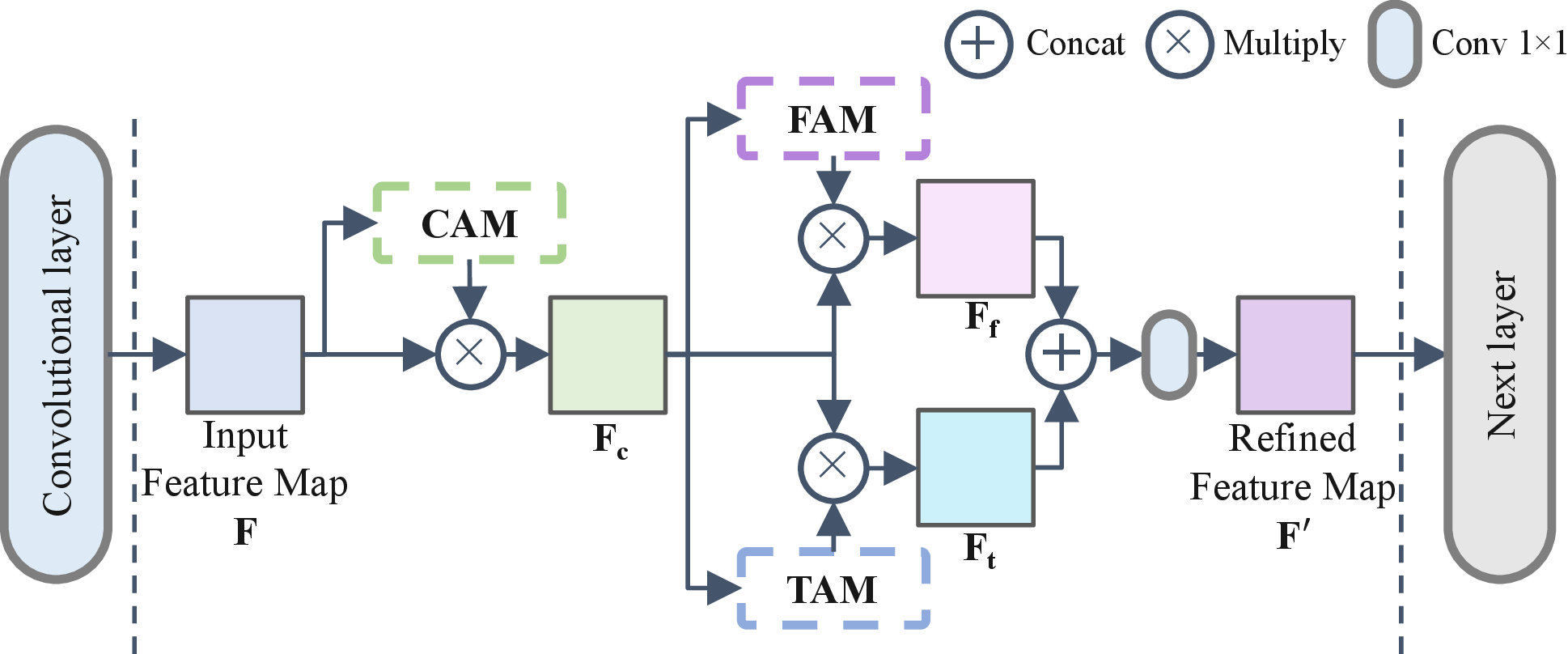}}
    \caption{Overview of the proposed TFA mechanism integrated in a convolutional layer.}
    \label{fig:overview}
\end{figure}

We propose a TFA mechanism for extracting meaningful channel, frequency, and time information of the spectrogram inputs for AMR. The proposed TFA aims at devoting more computing power to that small but important part of the data.
The overview of the proposed TFA is shown in Fig. \ref{fig:overview}. The feature map generated from a convolutional layer is the input feature map of TFA, later the refined feature map generated by TFA is the input of next layer. The TFA contains three submodules, namely channel attention module (CAM), frequency attention module (FAM), and time attention module (TAM). The CAM is used to exploit the inter-channel relationship of features, and extract general information regarding channel importance in the input feature map. The FAM focuses on where is the important frequency information of the channel attention refined feature map, and TAM focuses on where is the important time information of the channel attention refined feature map.

Given an input feature map $\mathbf{F}\in R^{H\times W \times C}$ from previous convolutional layer, where $H$ and $W$ denote the height and width, respectively, and $C$ denotes the number of channels. The TFA first uses CAM $\mathbf{M_c}$ to generate a channel refined feature map $\mathbf{F_c}\in R^{H\times W \times C}$.
Later, FAM $\mathbf{M_f}$ and TAM $\mathbf{M_t}$ are performed on $\mathbf{F_c}$ in parallel.
The parallel attention operations are performed to generate a frequency refined feature map $\mathbf{F_f} \in R^{H\times W \times C}$ and a time refined feature map $\mathbf{F_t} \in R^{H\times W \times C}$, respectively.
After concatenation of the two refined feature maps $\mathbf{F_f}$ and $\mathbf{F_t}$,
a convolution layer with 1$\times$1-sized kernel is applied to generate the final refined feature map $\mathbf{F'} \in R^{H\times W \times C}$. Then $\mathbf{F'}$ is treated as the input of the next layer.
The overall process can be summarized as:
\begin{gather}
    \mathbf{F_c} = \mathbf{M_c}(\mathbf{F} ) \otimes \mathbf{F}, \
    \mathbf{F_f} = \mathbf{M_f}(\mathbf{F_c} ) \otimes \mathbf{F_c},\
    \mathbf{F_t} = \mathbf{M_t}(\mathbf{F_c} ) \otimes \mathbf{F_c},\\
    \mathbf{F'}  =f^{1\times1}([\mathbf{F_f};\mathbf{F_t}]),
\end{gather}
\noindent where $\otimes$ denotes element-wise multiplication.
Multiplication process enhances the important parts of input data and fade out the rest according to the learned attention operations $\mathbf{M_c}$, $\mathbf{M_f}$ and $\mathbf{M_t}$.
$f^{1\times1}$ denotes a convolutional layer with 1$\times$1-sized kernel.

\begin{figure*}[htbp]
    \centering{\includegraphics[scale=0.22]{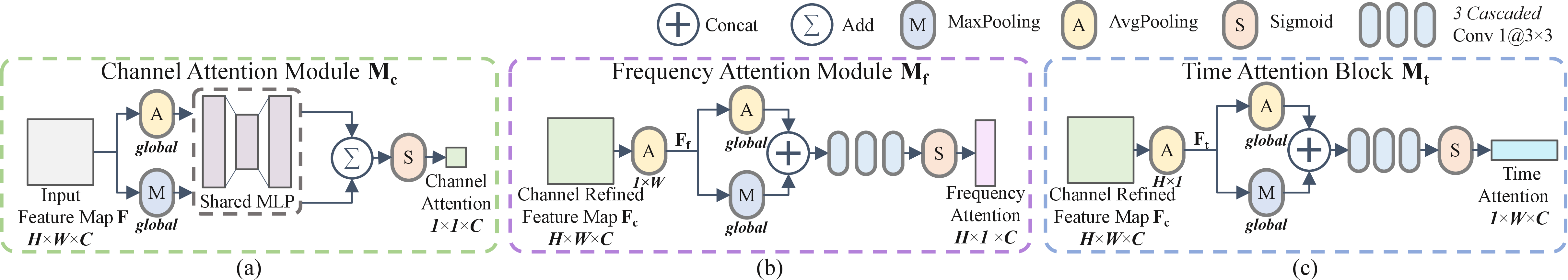}}
    \caption{Flowchart of the three submodules in TFA. (a) Channel attention module. (b) Frequency attention module. (c) Time attention module.}
    \label{fig:attention}
\end{figure*}

Fig. \ref{fig:attention} shows the procedures of CAM, FAM, and TAM. The CAM first performs global max-pooling and global average-pooling on the input feature map to generate features that denote two different contexts respectively. The features are then used as the input of a shared network, which consists of a multi-layer perceptron (MLP) comprising two densely connected layers with $C/8$ and $C$ neurons. The MLP is trained with the network with same training settings.
The outputs of the shared network are element-wise added up. Then a sigmoid function is performed to generate the channel attention map. The operation $\mathbf{M_c}$ is given as:
\begin{equation}
    \mathbf{M_c}(\mathbf{F}) = \sigma(MLP(AvgPool(\mathbf{F}))+MLP(MaxPool(\mathbf{F}))),
\end{equation}
\noindent where $\sigma$ denotes the sigmoid function. The output channel attention map $\mathbf{M_c}(\mathbf{F}) \in R^{1\times 1 \times C}$ is multiplied by $\mathbf{F}$ to generate a channel attention refined feature map $\mathbf{F_c}$.

The FAM and TAM have similar procedures. The TAM focuses on the x-axis of input spectrogram which represent the time axes, and FAM focuses on y-axis which represent the frequency axes. The channel refined feature map $\mathbf{F_c}$ is the input of FAM and TAM.
The FAM averages $\mathbf{F_c}$ along the time axes to focus on the frequency feature $\mathbf{F_f}\in R^{H\times 1 \times C}$, which is given by $\mathbf{F_f} = AvgPool_{1 \times W}(\mathbf{F_c})$. Then, max-pooling and average-pooling are performed and then concatenated to further extract the frequency feature. After that, three cascaded $3\times3$ convolutional layers and a sigmoid function are performed to generate a frequency attention map $\mathbf{M_f}(\mathbf{F_c}) \in R^{H\times 1 \times C}$. Unlike FAM, TAM averages $\mathbf{F_c}$ along the frequency axes to focus on the time feature $\mathbf{F_t} \in R^{1\times W \times C}$, which is given by $\mathbf{F_t} = AvgPool_{H \times 1}(\mathbf{F_c})$, and performs the same operations as in FAM to generate a time attention map $\mathbf{M_t}(\mathbf{F_c}) \in R^{1\times W \times C}$.
The operations $\mathbf{M_f}$ and $\mathbf{M_t}$ are given as:
\begin{equation}
    \begin{aligned}
        \mathbf{M_f}(\mathbf{F_c}) & = \sigma(f^{3 \times 3}_3([AvgPool(\mathbf{F_f}); MaxPool(\mathbf{F_f})])), \\
        \mathbf{M_t}(\mathbf{F_c}) & = \sigma(f^{3 \times 3}_3([AvgPool(\mathbf{F_t}); MaxPool(\mathbf{F_t})])),
    \end{aligned}
\end{equation}
\noindent where $f^{3 \times 3}_3$ denotes 3 cascaded convolutional layers with $3 \times 3$-sized kernel.
The frequency attention map $\mathbf{M_f}(\mathbf{F_c})$ and time attention map $\mathbf{M_t}(\mathbf{F_c})$ are multiplied by the input feature map $\mathbf{F_c}$ to generate a frequency refined feature map $\mathbf{F_f}$ and a time refined feature map $\mathbf{F_t}$, respectively. After concatenating $\mathbf{F_f}$ and $\mathbf{F_t}$, a convolutional layer $f^{1\times1}$ is used to generate the final refined feature map $\mathbf{F'}$ as shown in Fig. \ref{fig:overview}.

\subsection{Network Architecture}
\begin{figure}[htbp]
    \centering{\includegraphics[scale=0.26]{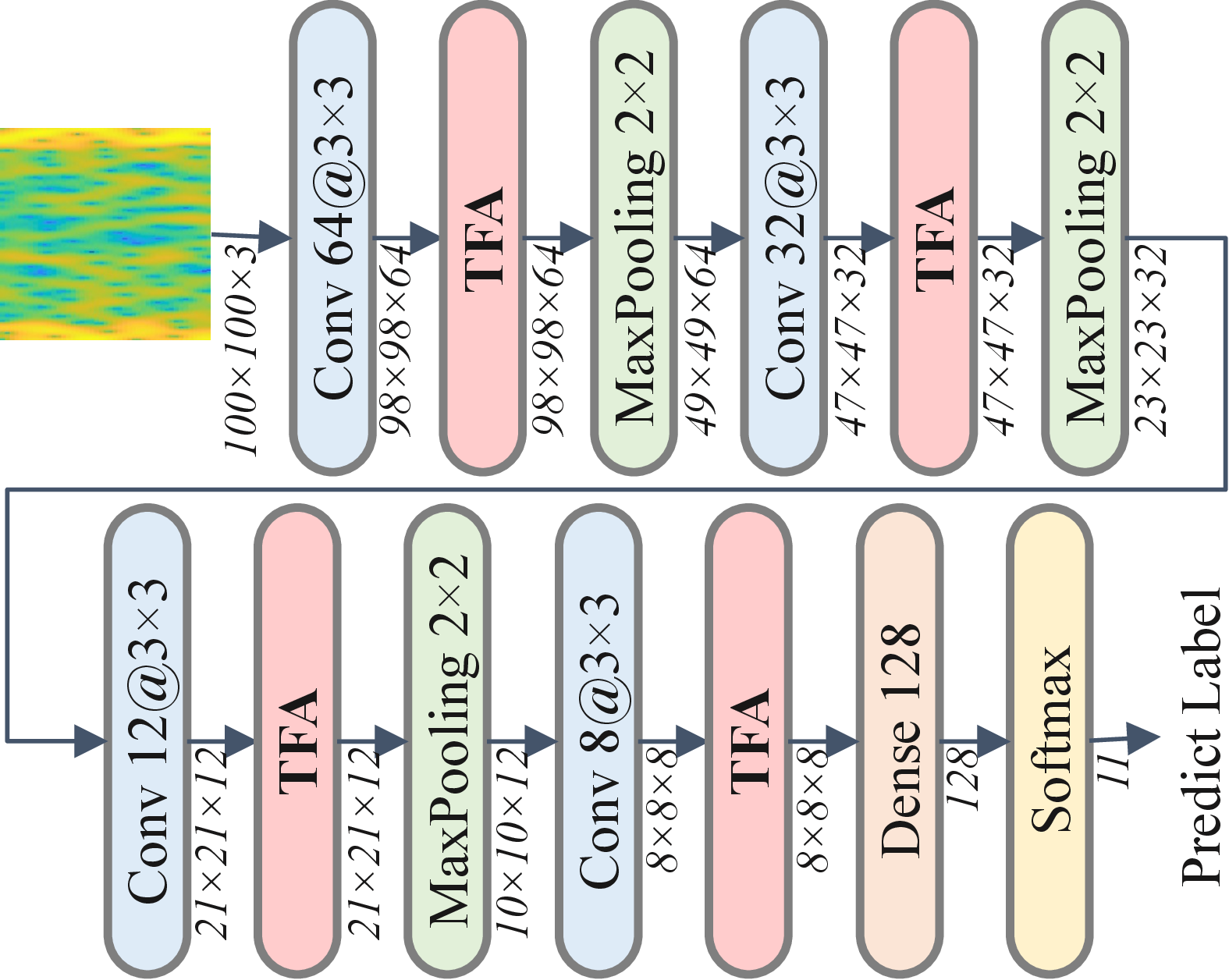}}
    \caption{Network architecture of the proposed TFA-SCNN.}
    \label{fig:cnn}
\end{figure}

We follow a CNN architecture similar to the one used in \cite{cite:SCNN}, called spectrum CNN (SCNN), and investigate the designed TFA block into the CNN architecture, called TFA-SCNN.
Fig. \ref{fig:cnn} illustrates the overview of the framework TFA-SCNN. It consists of one input layer, 4 convolutional layers integrated with TFA, one densely connected layer, and an output softmax layer.
The input of the network is a spectrogram image with the dimension of $100\times100\times3$. The convolutional layers use $3\times3$-sized kernel and the number of kernels of the 4 convolutional layers is setting as 64, 32, 12, 8. The feature maps from convolutional layers integrated with TFA are followed by rectified linear unit (ReLU) \cite{cite:ReLU} and a max-pooling layer with a size of $2\times2$, except for the last one only followed by ReLU.
Specifically, the dimension of the feature maps generated by TFA is $98\times98\times64$, $47\times47\times32$, $21\times21\times12$, $8\times8\times8$.
The densely connected layer consists of 128 neurons. The output of the network is the predicted modulation mode of input. The network is trained using stochastic gradient descent to minimize the cross-entropy loss function, that is, $\mathbf{w}^{*}=\operatorname{argmin}_{\mathbf{W}} \frac{1}{N} \sum_{i=1}^{N}\mathcal{L}(\mathbf{w}; x^i, t^i)$, with the number of training examples $N$, the true labels $t^i$, and the predicted labels $x^i$. $\mathcal{L}$ denotes the loss function, that is, $\mathcal{L}=-\sum_{j}^{M} \beta_j log(q_j)$, where $M$ denotes the number of classes, $\beta_j$ is a binary indicator with $\beta_j = 1$ if $x^i$ is $t^i$, otherwise $\beta_j = 0$, and $q_j$ denotes the predicted probability of belonging to class $j$.

\section{Experiments}
\subsection{Dataset}
We evaluate the proposed AMR framework on an open-source dataset RadioML2016.10a \cite{cite:radioML} and its larger version RadioML2016.10b.
RadioML2016.10a consists of analog and digital modulation methods, including 11 commonly used modulations modes in communication systems, which are 8PSK, AM-DSB, AM-SSB, BPSK, CPFSK, GFSK, PAM4, QAM16, QAM64, QPSK, and WBFM. RadioML2016.10a includes 220000 modulated signals with 20 different signal-to-noise ratios (SNRs) ranging from $-20$ dB to $18$ dB, and 1000 signals per SNR per modulation mode.
RadioML2016.10b consists of 1200000 modulated signals with 20 different SNRs ranging from -20 dB to 18 dB and 6000 signals per SNR per mode. Each signal in the both datasets consists of complex IQ.
Unlike most deep learning-based AMR methods, where IQ information is directly used as two-dimensional signals, we generate one dimensional complex signal using the given IQ information. To train, validate and test the learning-based AMR models, in our experiments, for each modulation mode per SNR, we randomly split the dataset into training set, validation set and test set, the corresponding ratio is 7:1:2.

\subsection{Experiment Setup}
We use three experiments to analyze the effectiveness of the proposed TFA mechanism and the recognition performance of the proposed models. First, we study the effect of different combination modes of the frequency and time attention mechanisms on recognition accuracy. Second, using the proposed model SCNN with no attention as a baseline, we evaluate the performance of the proposed TFA on SCNN, called TFA-SCNN. After that, we evaluate the performance of the proposed model additionally with Gaussian filter-based noise reduction \cite{cite:SCNN}, called TFA-SCNN2, and compare the proposed TFA-SCNN and TFA-SCNN2 with the existing SOTAs  with open source code:
\begin{itemize}
    \item IQ-CNN \cite{cite:IQCNN}: a CNN-based method, which uses IQ information as the input.
    \item CLDNN \cite{cite:CLDNN}: a convolutional long short-term deep neural networks (CLDNN)-based method with optimal parameters for AMR.
    \item AAM-SCNN \cite{cite:liang2021automatic}: a baseline model with an adaptive attention mechanism module (AAM).
\end{itemize}

For signal representation, we convert the complex signals into spectrogram images using frame-based STFT, with a 95\% overlapping Hamming window and a frame length of 40 samples. The resolution of the input spectrograms is $100 \times 100 \times 3$. We normalize all spectrograms before processing, and use root-mean-square prop (RMSprop) as the optimizer. The learning rate starts with 0.0005 and is reduced by a factor of 0.1 when validation loss does not drop within 10 epochs. The training process is terminated when validation loss does not drop within 15 epochs, and the model with the smallest validation loss is saved and used for testing. All experiments are implemented using Keras with Tensorflow backbone and NVIDIA RTX 3090 GPU platform.

\subsection{Experiment Results}
Table \ref{table:ablation} shows the results of the ablation experiments on baseline model SCNN. The TFA outperforms all other variants by a significant performance improvement. Whether cascaded CAM-FAM or cascaded CAM-TAM, the recognition accuracy is improved compared to SCNN without attention, which indicates that the attention mechanism extracting meaningful frequency or time features can improve recognition accuracy. In addition, experiment results in Table \ref{table:ablation} show that cascaded FAM and TAM performs worse than the proposed parallel architecture in TFA.
The ablation study shows that simultaneous modelling of frequency and time importance from spectrograms in CNN improves recognition accuracy.
\begin{table}[htbp]
    \renewcommand{\arraystretch}{1.3}
    \caption{Ablation Experiment Results on RadioML2016.10a}
    \begin{center}
        \begin{tabular}{c|c|c|c|c}
            \hline \hline
            \multirow{2}{*}{Attention Variant} & \multicolumn{4}{c}{Accuracy}                                                    \\
            \cline{2-5}
                                               & $-8$ dB                       & $-2$ dB         & $4$ dB          & $10$ dB         \\
            \hline
            None                               & 0.372                        & 0.687          & 0.801          & 0.823          \\
            \hline
            cascaded CAM-FAM                   & 0.395                        & 0.732          & 0.818          & 0.841          \\
            \hline
            cascaded CAM-TAM                   & 0.396                        & 0.740          & 0.818          & 0.834          \\
            \hline
            cascaded CAM-TAM-FAM               & 0.389                        & 0.729          & 0.811          & 0.843          \\
            \hline
            \textbf{proposed}                  & \textbf{0.426}               & \textbf{0.766} & \textbf{0.864} & \textbf{0.857} \\
            \hline \hline
        \end{tabular}
        \label{table:ablation}
    \end{center}
\end{table}

\begin{figure}[htbp]
    \centering{\includegraphics[scale=0.21]{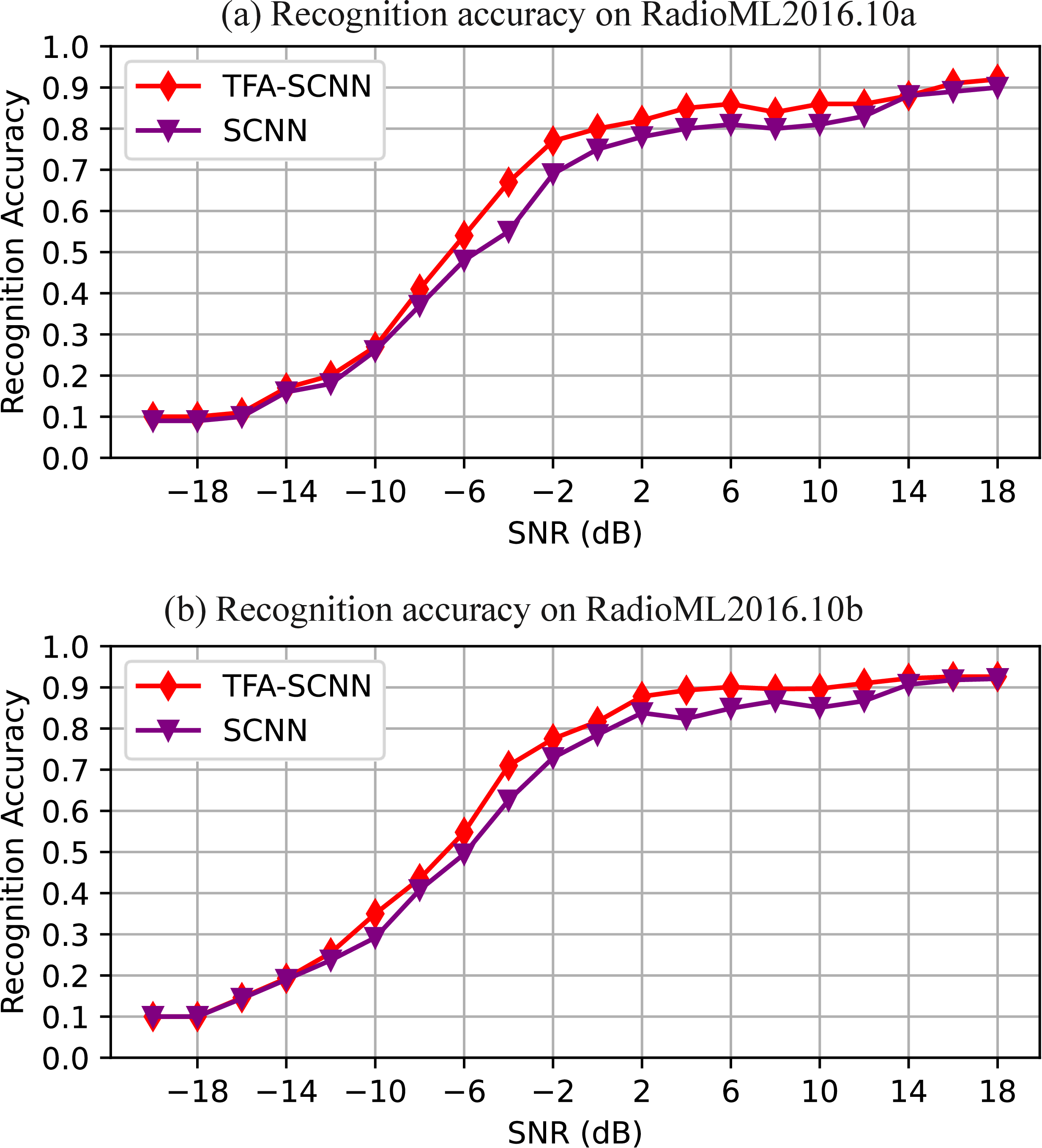}}
    \caption{Recognition accuracy of TFA-SCNN and SCNN.}
    \label{fig:resultSCNNab}
\end{figure}

Fig. \ref{fig:resultSCNNab} shows the recognition accuracy comparison between SCNN and TFA-SCNN versus SNR on RadioML2016.10a and RadioML2016.10b.
Compared to the baseline model SCNN, the proposed TFA-SCNN has higher recognition accuracy. Specifically, on RadioML2016.10a, the recognition accuracy of TFA-SCNN is around 2\% to 4\% higher than those of the SCNN when SNR is above $10$ dB, and around 5\% to 8\% higher than those of SCNN when SNR is between $-8$ dB and $10$ dB.
On dataset RadioML2016.10b, the recognition accuracy of TFA-SCNN is around 1\% to 5\% higher than those of the SCNN when SNR is above $10$ dB. The TFA-SCNN gets around 3\% to 9\% higher accuracy than SCNN when SNR is between $-8$ dB and $10$ dB.

\begin{figure}[htbp]
    \centering{\includegraphics[scale=0.10]{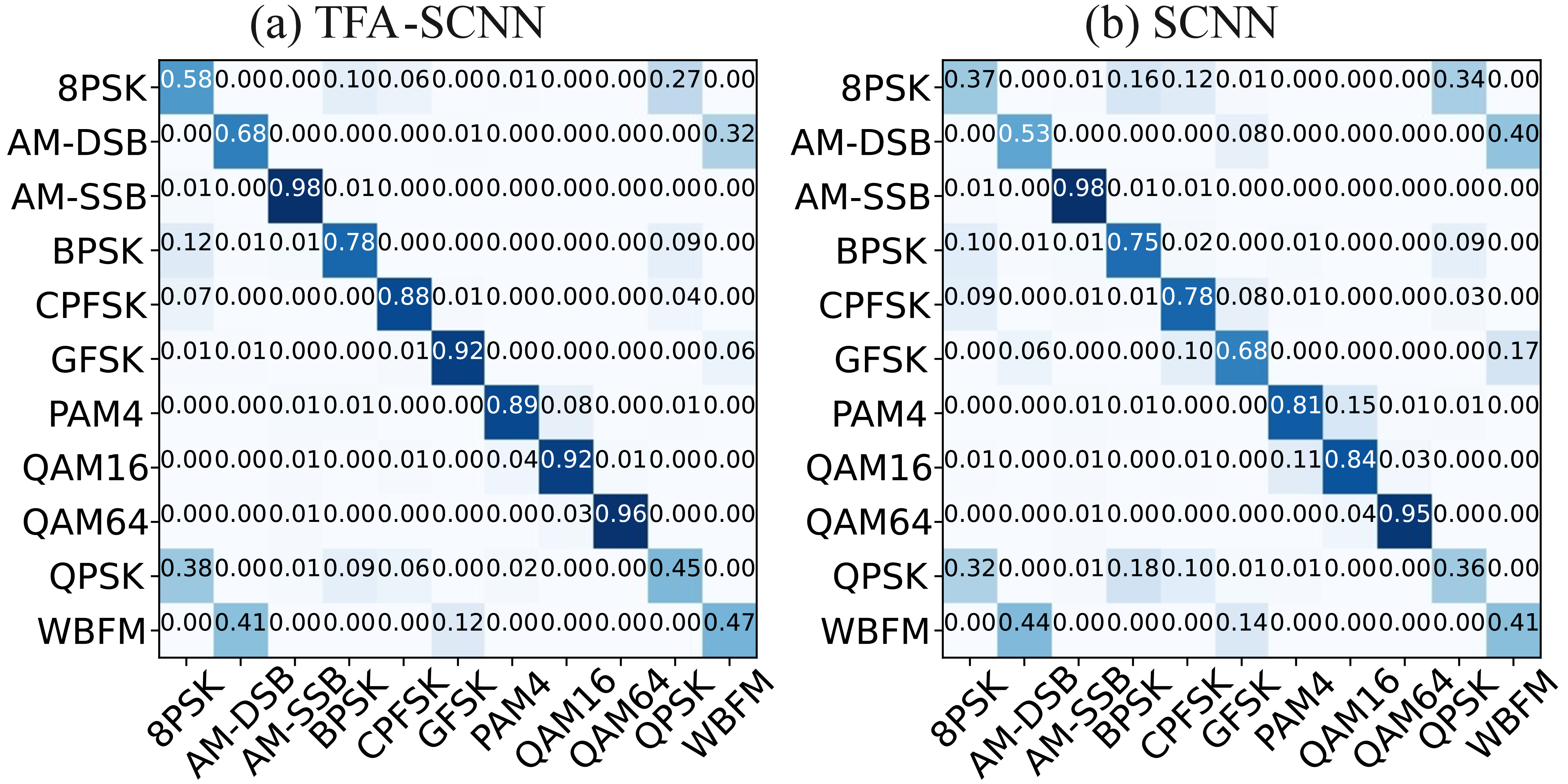}}
    \caption{Confusion matrices on RadioML2016.10a at $-2$ dB SNR.}
    \label{fig:confMat}
\end{figure}

Fig. \ref{fig:confMat} shows confusion matrices of TFA-SCNN and SCNN at $-2$ dB SNR on RadioML2016.10a. The results show that TFA is able to improve the recognition accuracy of all modulation modes, especially for modes: 8PSK, AM-DSB, and GFSK, getting around 15\% to 24\% performance improvement.
The confusion problem between WBFM and AM-DSB is because that both of them belong to analog modulation, and the signal data were generated using the same audio source signal with silent segments, making some of their spectrogram features more difficult to distinguish. Another confusion problem is between 8PSK and QPSK, since the main difference between 8PSK and QPSK is in phase, while spectrograms are weak in representing phase information.

\begin{figure}[htbp]
    \centering{\includegraphics[scale=0.72]{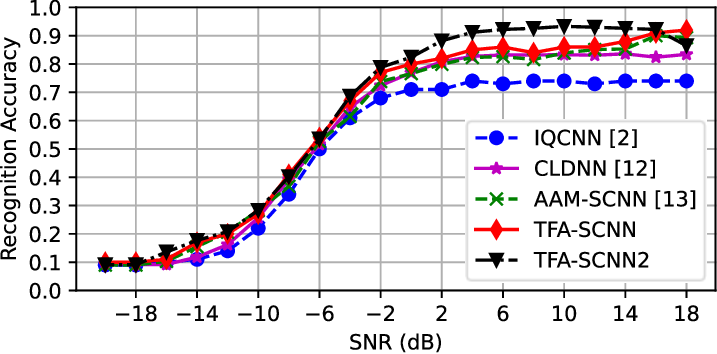}}
    \caption{Recognition accuracy comparison on RadioML2016.10a.}
    \label{fig:result}
\end{figure}

Next, we compare the recognition accuracy of TFA-SCNN and TFA-SCNN2 with IQCNN, CLDNN, and AAM-SCNN on RadioML2016.10a. The experiment results are shown in Fig. \ref{fig:result}.
We observe that the presented models with TFA (TFA-SCNN and TFA-SCNN2) perform better than the other methods when SNR is above $-14$ dB.
Specifically, TFA-SNN2 performs clearly better than the other methods when SNR is above $2$ dB, but the accuracy gets around 3\% to 4\% lower than TFA-SCNN and AAM-SCNN at $18$ dB SNR. This is consistent with the results of SCNN and SCNN2 in literature \cite{cite:SCNN}, since the noise reduction algorithm has limited capability to improve recognition accuracy when signals are severely distorted and close to clean.
In addition, the accuracy of TFA-SCNN gets around 1\% to 3\% higher than that of AAM-SCNN at all SNR levels, and it achieves a maximum recognition accuracy of 92\% at $18$ dB SNR. This can be explained that the TFA mechanism considers the inherent characters of the time-frequency analysis and extracts important information in terms of channel, frequency and time dimensions, while the AAM mechanism pays attention to channel and spatial information.

\begin{table}[htbp]
    \renewcommand{\arraystretch}{1.3}
    \caption{Computational Complexity Comparison}
    \begin{center}
        \begin{tabular}{c|c|c|c}
            \hline \hline
            Model    & Training Time & Inference Time & Parameters \\
            \hline
            IQCNN    & 0.0383ms      & 0.0380ms       & 2830k      \\
            \hline
            CLDNN    & 0.0477ms      & 0.0382ms       & 167k       \\
            \hline
            AAM-SCNN & 0.6163ms      & 0.0395ms       & 94k        \\
            \hline
            SCNN     & 0.2071ms      & 0.0358ms       & 92k        \\
            \hline
            TFA-SCNN & 1.6798ms      & 0.0391ms       & 104k       \\
            \hline \hline
        \end{tabular}
        \label{table:computation}
    \end{center}
\end{table}
Furthermore, we compare computational complexity between TFA-SCNN and SOTAs in terms of the average training time, the average inference time and the amount of learned parameters. The comparison results are shown in Table \ref{table:computation}.
TFA-SCNN with TFA block costs much more training time (around 1.2ms) compared to the baseline model SCNN, but the inference time of TFA-SCNN increases very little (around 0.003ms).
TFA-SCNN has slightly more model parameters than SCNN, but fewer parameters than IQCNN and CLDNN.

\section{Conclusion}
In this work, we presented a CNN-based framework for automatic modulation recognition with a novel TFA mechanism. The TFA is performed on input feature maps to generate attention refined feature maps by learning feature representations for explicitly attending to important channel, frequency, and time information. Experiment results demonstrated the effectiveness of modelling TFA in the CNN front-end, and the presented CNN models with TFA (TFA-SCNN and TFA-SCNN2) outperform three existing learning-based methods from literature. The proposed attention mechanism causes additional computational burden than the baseline model SCNN, but requires similar inference time as the other methods and less learned parameters than IQ-CNN and CLDNN. The performance improvement of the proposed frameworks is incremental in the presence of complex channel environment and low SNRs. As a future work, we plan to extend the method to improve recognition performance at low SNRs using deep learning-based signal enhancement techniques.

\bibliographystyle{IEEEtran}
\bibliography{Gong_WCL2021-2046.R1}

\end{document}